\documentclass[aps,prl,superscriptaddress,twocolumn,showpacs,floatfix]{revtex4}
\usepackage{epsfig}

\newcommand{\eg}{{\it e.g.\ }}
\newcommand{\ie}{{\it i.e.\ }}
\newcommand{\ignore}[1]{}

\begin{document}
\title{(Un)detectable cluster structure in sparse networks}
\author{J\"org Reichardt}
\affiliation{Institute for Theoretical Physics, University of W\"urzburg, 
97074 W\"urzburg, Germany}
\author{Michele Leone}
\affiliation{ISI Foundation, Viale S. Severo 65,I-10133 Torino, Italy}
\date{\today}
\begin{abstract}
We study the problem of recovering a known cluster structure in a sparse network, also known as the planted partitioning problem, by means of statistical mechanics. We find a sharp transition from un-recoverable to recoverable structure as a function of the \emph{separation} of the clusters. For multivariate data, such transitions have been observed frequently, but always as a function of the \emph{number of data points} provided, \ie given a large enough data set, two point clouds can always be recognized as different clusters, as long as their separation is non-zero. In contrast, for the sparse networks studied here, a cluster structure remains undetectable even in an infinitely large network if a critical separation is not exceeded. We give analytic formulas for this critical separation as a function of the degree distribution of the network and calculate the shape of the recoverability-transition. Our findings have implications for unsupervised learning and data-mining in relational data bases and provide bounds on the achievable performance of graph clustering algorithms. 
\end{abstract}
\pacs{89.75.Hc,89.65.-s,05.50.+q,64.60.Cn} \maketitle 

\noindent In any branch of science, exploratory data analysis often starts with clustering. Supposing the data are clustered, a natural question is whether one can at all hope to recover the underlying structure of the data from a finite number of samples \cite{EngelBuch}. This question has been studied extensively by Physicists in the following setting: Given $\alpha N$ data points from a known probability distribution forming clusters in an $N$ dimensional space (\eg a mixture of Gaussians), can we recover the parameters of the probability distribution from the given data alone and can we label the data points correctly as belonging to one of the point clouds? The answer that has been given is generally yes. For any non-zero separation of the clusters in a multidimensional space, one can learn the parameters of the underlying distribution if only the number of data points is large enough, \ie one observes a transition from unrecoverable to recoverable structure as a function of $\alpha$ \cite{Biehl93,VanDenBroeck96,Reimann96,Buhot98}. 

In this contribution, we study the problem of recovering a known cluster structure in networks. This problem has received considerable attention recently under the term ``community detection'' \cite{Girvan,GuimeraMeta,GuileraReview,ReichardtPRL} in the research on the topology of complex networks. If the clusters are all of equal size, this problem is known as the ``planted partition problem'' in computer science where the most studied case is the planted bisection problem in which all the nodes in the network are members of one of only two clusters. Any \emph{pair of  nodes} from the same cluster is connected with probability $p$, while any pair of nodes from distinct clusters is connected with probability $r<p$. As an example, Onsj\"o and Watanabe provide an algorithm that provably recovers the planted solution with probability $>1-\delta$ if $p-r=\Omega(N^{-1/2}\log(N/\delta))$ \cite{Onsjo2006}. Other authors present different algorithms with similar bounds \cite{Condon1999,Carson2001}. Again, if the data set, \ie the number of nodes $N$, is large enough, the two clusters can be recovered regardless of the strength of their separation $(p-r)$. The above bound applies only to dense networks in which the average number of connections per node $\langle k\rangle$ grows linearly with the number of nodes in the network. However, most real world networks on which clustering is performed are sparse and have link densities of the order of $1/N$ in which case the above bound is meaningless \cite{NewmanReview,BAReview}.  The mean connectivity of sparse networks does not grow as the system size. Consider for example the world wide web: doubling the number of web pages will not lead to doubling the number of links a single page lists or receives on average. We will show that for sparse networks, a pre-defined cluster structure remains unrecoverable even in the limit of infinite $N$, if the probability for an intra-cluster link does not exceed a critical value $p_{in}^c$ which depends on the degree distribution $p(k)$. We will calculate $p_{in}^c$ and the shape of the transition analytically as a function of $p(k)$.

Specifically, we consider the problem of recovering the pre-defined cluster structure in infinitely large sparse networks for which a degree distribution $p(k)$ is given and the same in all clusters. The average connectivity per node $\langle k\rangle$ is assumed to be finite. We parameterize the planted cluster structure of the network by the number of clusters $q$ and  the probability $p_{in}$ that a given \emph{edge} lies within one of $q$ equal sized clusters. Every node $i$ carries an index $s_i\in \{1,...,q\}$ indicating the cluster to which it belongs by design. For $p_{in}=1/q$ the pre-defined cluster structure cannot be recovered by definition, while for $p_{in}=1$ recovery is trivial as our network consists of $q$ disconnected parts. Given such a network, we are interested in finding a partition, \ie an assignment of indices $\sigma_i\in\{1,...,q\}$ to the $N$ nodes of the network, such as to maximize the accuracy $A=\sum_i\delta_{s_i,\sigma_i}/N$ of recovering the planted solution. Since we cannot assume knowledge of $p_{in}$, the best possible approach is to find a partition that minimizes the number of edges running between different parts, \ie a minimum cut-partition. Na\"{\i}vely, one would expect the overlap of the minimum cut partition with the planted solution and hence the accuracy to increase steadily with $p_{in}$ between $1/q$ and $1$. However, we will show that for sparse networks, the minimum cut partition is uncorrelated with the planted partition until $p_{in}$ exceeds some critical value $p_{in}^c$ which depends on $p(k)$. Hence, below $p_{in}^c$, the planted solution is impossible to recover. We will calculate $p_{in}^c$ and the maximum achievable accuracy as as a function of $p_{in}$ and $p(k)$. 

Let us formulate the problem of finding a minimum cut partition as finding the ground state of the following ferromagnetic Potts Hamiltonian:
\begin{equation}
\mathcal{H}_{Part}=-\sum_{i<j} J_{ij} \delta_{\sigma_i,\sigma_j} + \mbox{Constraint}.
\label{HPart}
\end{equation}
Here, $J_{ij}$ is the $\{0,1\}$ adjacency matrix of the graph and the constraint enforces a zero-magnetization ground state corresponding to an equi-partition. For graphs without cluster structure, \ie $p_{in}=1/q$, this problem has been studied extensively for Poissonian degree distributions or Bethe lattices with a fixed valence \cite{FuAnd,KanterSompQPart,Oliveira89,LaiGold,LiaoPRL}. A recent result generalizes to arbitrary degree distributions \cite{ReichardtPRER}. Note that the energy of the planted partition is $E^p=p_{in}\langle k\rangle/2$ with $\langle k\rangle$. To study the ground state of (\ref{HPart}) we employ the Bethe-Peirls approach from statistical mechanics, also known as the cavity method or belief propagation,  directly at zero temperature \cite{CavZeroT}. At an informal level, for the ferrogmagntic system studied here, this method can be described as the following process: nodes are assumed to pass messages ${\bf u}$ among each other across the links of the network. A message from node $i$ to $j$ is a $q$-dimensional vector of zeros and ones. An entry of one in component $s$ of ${\bf u}$ indicates to node $j$ that node $i$ would like node $j$ to assume state $\sigma_j=s$. To generate this message to $j$, node $i$ has taken all messages coming from all other nodes $k\neq j$ connected to $i$ and summed them to obtain a so-called cavity field ${\bf h}_{i\to j}=\sum_{k\neq j}J_{ki}{\bf u}_{k\to i}$. Then, node $i$ converts this cavity field into a message to $j$ via ${\bf u}_{i\to j}={\bf \hat{u}}({\bf h}_{i\to j})$. In our case, the function ${\bf \hat{u}}$ is defined via
\begin{eqnarray}
v({\bf h}) & = & \max(h^1,...,h^q),\\
\hat{u}^s({\bf h}) & = & \max(h^1,h^s+1,...,h^q)-v({\bf h}).
\label{WandU}
\end{eqnarray}
This means that ${\bf \hat{u}}$ picks the maximum components in ${\bf h}$ and sets all corresponding components in ${\bf u}$ to one and the rest to zero. Due to possible degeneracy in the components of ${\bf h}$, the vector ${\bf u}={\bf \hat{u}}({\bf h})$ may have more than one non-zero entry and is never completely zero. This observation is fundamental for all further developments. The field components of ${\bf h}$ take only integer values, because we only have integer couplings $J_{ij}$ between the spins.
This process of message passing is iterated until a stationary state is reached corresponding to the replica symmetric ground state. It is fully described by the probability distribution $Q^s({\bf u})$ of messages being sent in the system. The superscript $s$ denotes a possible dependence of this distribution on the index of the pre-defined cluster to which the sending node belongs. An easy to follow formal derivation of a set of self-consistent integral equations for $Q^s({\bf u})$ can be found in Refs. \cite{Braunstein,WeigtCorr}. It is general in that the particular form of the Hamiltonian enters only via the functions $v({\bf h})$ and ${\bf \hat{u}}({\bf h})$ and is therefore not repeated here. 

There are $2^q-1$ possible messages ${\bf u}$. Since the probabilities of sending them may depend on the planted cluster from which they are sent, there are in principle $q(2^q-1)$ different probabilities $Q^s({\bf u})$ to determine. We are only interested in distributions that allow to fulfill the constraint of an equi-partition and that are symmetric under permutation of the indices as is our planted cluster structure. These conditions reduce the number of different probabilities $Q^s({\bf u})$ to only $2q-1$ parameters $\eta_{cw}$:
 \begin{equation}
Q^s({\bf u})=\eta_{cw}\mbox{, where } c=u^s \mbox{ and } w=||{\bf u}||^2-c.
\end{equation}
Here, $u^s$ denotes the $s^{th}$ component of the message vector ${\bf u}$ under consideration. Without loss of generality, we have thus implicitly introduced a preferred direction for each planted cluster. The probability $Q^s({\bf u})$ that a node from planted cluster $s$ sends a message ${\bf u}$ depends only on whether or not ${\bf u}$ has an entry of one in the ``correct'' component $s$ ($c=1$) and on how many ``wrong'' components $w$ in ${\bf u}$ carry an entry of one ($w\in\{1-c,..,q-1\}$). It is understood that for $p_{in}\to 1$ we have $\eta_{10}\to 1$, \ie only ``correct'' messages are sent. Equivalently, for $p_{in}\to 1/q$ we must have $\eta_{1,\alpha-1}=\eta_{0,\alpha}=\eta_{\tau}$, \ie the probability of a message depends only on the number $\tau=w+c$ of non-zero entries in it.
The $2q-1$ new order parameters $\eta_{cw}$ which describe  $Q^s({\bf u})$ obey the following normalization:
\begin{equation}
\sum_{c=0}^1\sum_{w=1-c}^{q-1}\left(\begin{array}{c}q-1\\ w\end{array}\right)\eta_{cw}=1.
\end{equation}

\begin{figure*}[t] 
\includegraphics[width=5.75cm]{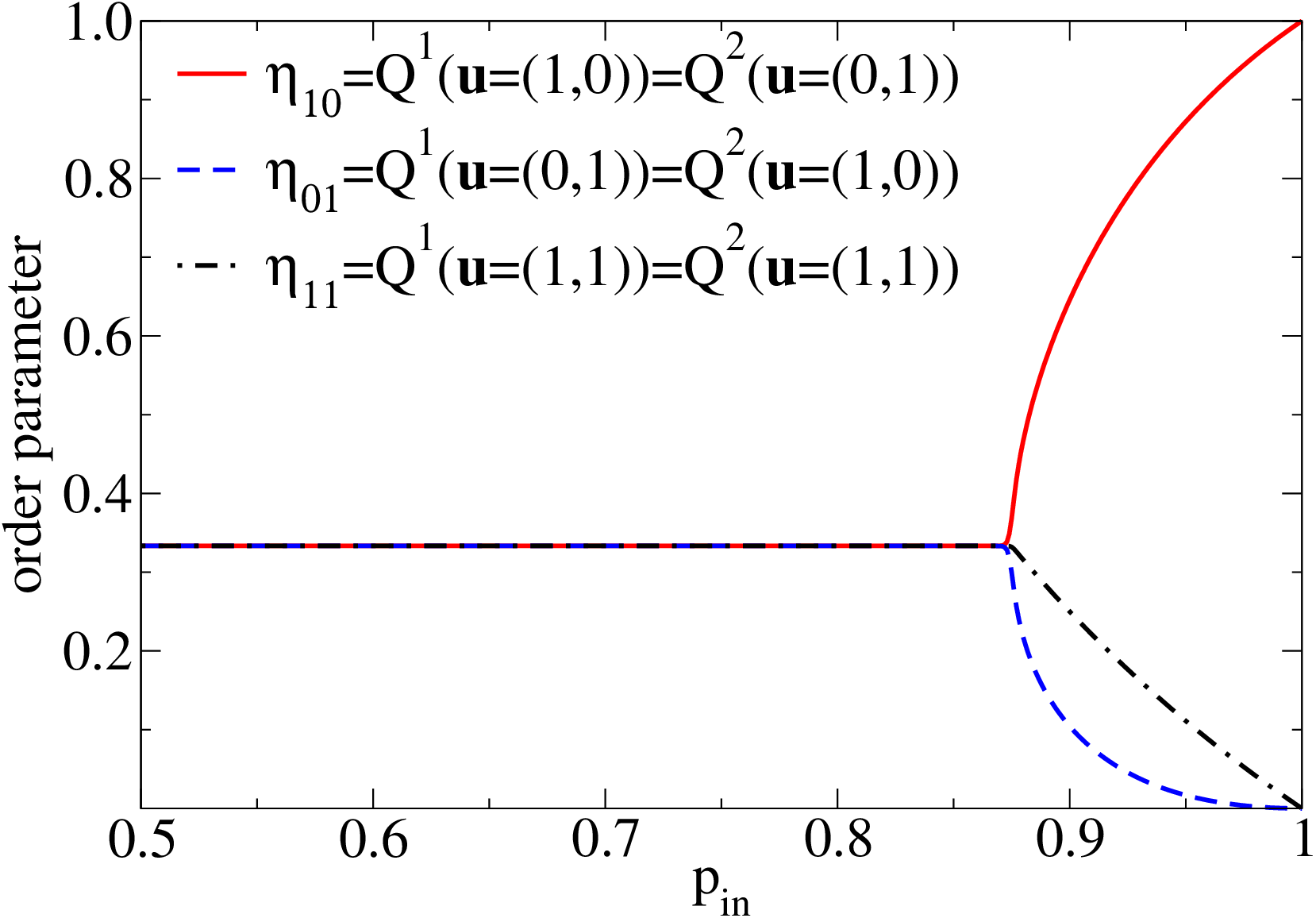}
\includegraphics[width=5.75cm]{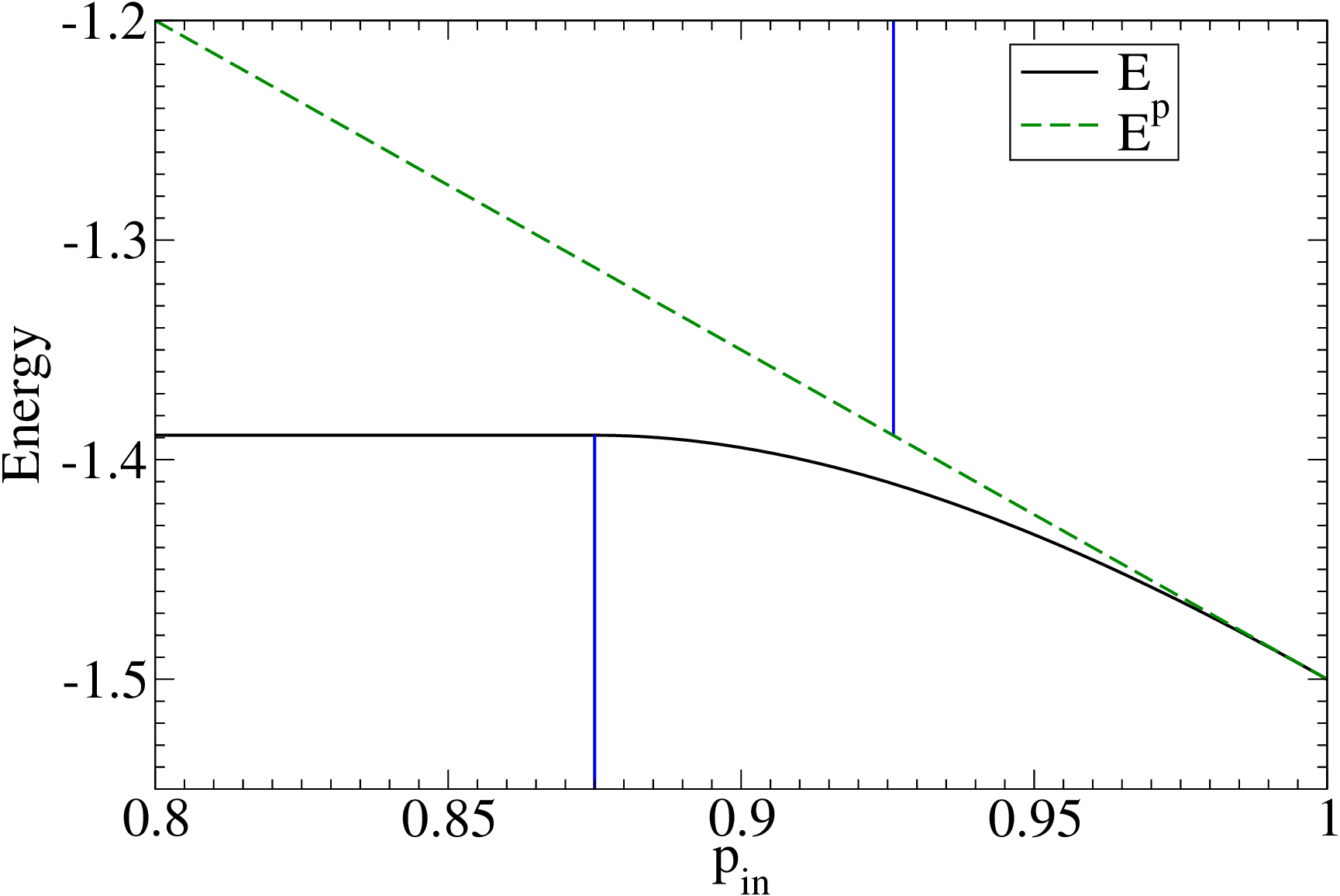}
\includegraphics[width=5.75cm]{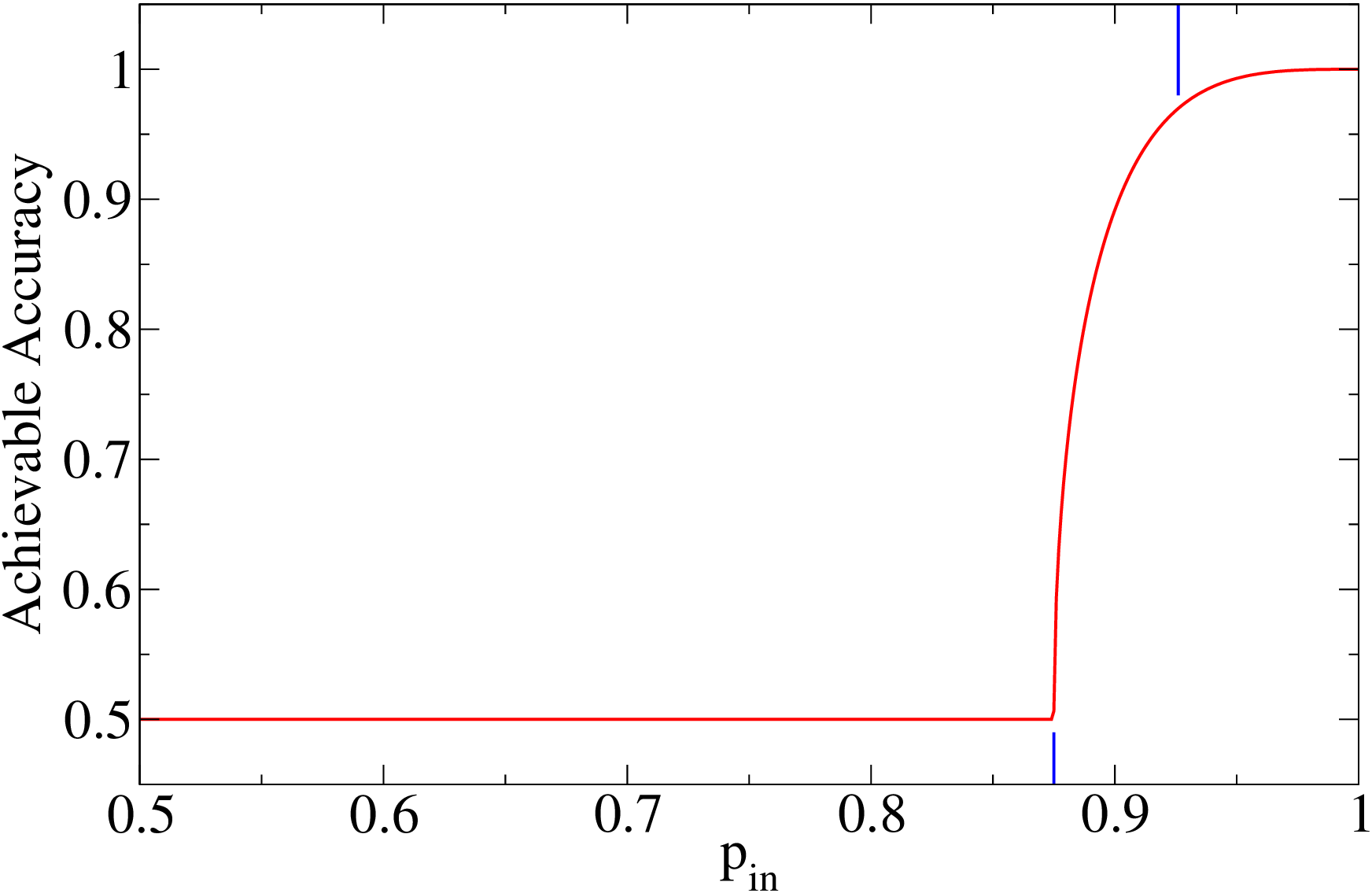}
\caption{\textbf{Left: }The order parameters $\eta_{cw}$ for the planted bisection problem on a random Bethe lattice with $k=3$ links per node as a function of $p_{in}$. The planted cluster structure in the network does not influence the ground state configuration until a critical value of $p_{in}$ is reached. \textbf{Middle: }The ground state energy $E$ of (\ref{HPart}) and the energy of the planted cluster structure $E^p$ vs. $p_{in}$. The left vertical blue line indicates the critical value of $p_{in}^c$ beyond which $\eta_{10}>\eta_{01}$ and $E<E^{Rnd}$ and the planted cluster structure starts to influence the ground state energy. The right vertical blue line indicates the na\"{\i}ve value of $p_{in}^n=2E^{Rnd}/\langle k\rangle$ beyond which $E^p<E^{Rnd}$. \textbf{Right: }The accuracy with which the planted cluster structure may be recovered. Again, the two vertical lines indicate $p_{in}^c$ and $p_{in}^n$.}
\label{Bethek3q2OrderParam}
\end{figure*}

Let us now turn to the problem of a planted bisection, \ie the case of two clusters. Then, we only have three possible messages ${\bf u}\in\{(1,0),(0,1),(1,1)\}$ and three order parameters $\eta_{cw}$. The self consistent integral equation for $Q^s({\bf u})$ can then be written as a set of polynomial equations for the $\eta_{cw}$ in a simple way:
\begin{small}
\begin{eqnarray}
\eta_{11}& = &\sum_{n0=0}^\infty \sum_{n=0}^\infty q(n_0+2n)\frac{(n_0+2n)!}{n_0!n!n!}\left(\eta_{10}^{in}\right)^n\left(\eta_{01}^{in}\right)^n\eta_{11}^{n_0}\mbox{\hspace{2mm}}\label{Eta11Full}\\
\eta_{10} & = &\sum_{n0=0}^\infty \sum_{n_1>n_2}^\infty q(n_0+n_1+n_2)\frac{(n_0+n_1+n_2)!}{n_0!n_1!n_2!}\times\nonumber\\
& & \times\left(\eta_{10}^{in}\right)^{n_1}\left(\eta_{01}^{in}\right)^{n_2}\eta_{11}^{n_0}\label{Eta10Full}
\end{eqnarray}
\end{small}
Together with the normalization condition $1=\eta_{10}+\eta_{01}+\eta_{11}$ this formes a closed set of equations in which $q(d)=(d+1)p(d+1)/\langle k\rangle$ denotes the excess degree distribution and  we have used the abbreviations $\eta_{10}^{in}=p_{in}\eta_{10}+(1-p_{in})\eta_{01}$ and $\eta_{01}^{in}=p_{in}\eta_{01}+(1-p_{in})\eta_{10}$. Equations (\ref{Eta11Full}-\ref{Eta10Full}) are easily solved for any value of $p_{in}$ and any degree distribution $p(k)$ by iteration. We see for $p_{in}=1/2$, we must have $\eta_{10}=\eta_{01}=\eta_1$ and only one independent order parameter remains. 

We assume node $i$ is assigned state $\sigma_i$ corresponding to the maximum component of the effective field ${\bf h}_{\mbox{\tiny eff}}=\sum_j J_{ji}{\bf u}_{j\to i}$.  In case of degeneracy, $\sigma_i$ is chosen with equal probability among the different maximum components. Given the distribution $Q^s({\bf u})$, we can thus calculate the probability that a node is assigned into correct pre-defined cluster $p(\sigma_i=s|s)$  from which the accuracy follows. The ground state energy of the partitioning problem is then given by:
\begin{small}
\begin{equation}
E  =  -\frac{\langle k\rangle}{2}\left(1+2(X-\eta_{10}\eta_{01})-(1-p_{in})(\eta_{10}-\eta_{01})^2\right),
\label{BiPartEnergy}
\end{equation}
\end{small}
where we have introduced $X$ as an abbreviation for 
\begin{small}
\begin{equation}
X  =  \frac{1}{\langle k\rangle}\sum_{n_0=0}^{\infty}\sum_{n=1}^{\infty}p(n_0+2n)\frac{(n_0+2n)!}{n_0!n!(n-1)!}\eta_{11}^{n_0}\left(\eta_{10}^{in}\right)^n\left(\eta_{01}^{in}\right)^n.\label{FunctionX}
\end{equation}
\end{small}
In case of a Poissonian degree distribution $p(k)=e^{-\lambda}\lambda^k/k!$ with mean $\lambda$, we can express this using Modified Bessel Functions of the first kind $I_1(n,x)$:
\begin{equation}
X_\lambda= \sqrt{\eta_{10}^{in}\eta_{01}^{in}}e^{-\lambda(1-\eta_{11})}I_{1}\left(1,2\lambda\sqrt{\eta_{10}^{in}\eta_{01}^{in}}\right).
\label{XPoisson}
\end{equation}
Let us denote with $E^{Rnd}$ the ground state energy for $p_{in}=1/q$.

Figure \ref{Bethek3q2OrderParam} shows the order parameters, the ground state energy and the achievable accuracy of recovering the planted bisection as a function of $p_{in}$ for a random Bethe lattice with exactly three neighbors per node. First, the order parameters $\eta_{10}$ and $\eta_{01}$, \ie the probabilities of sending a message indicating the correct or wrong cluster, respectively, are equal until a critical value of $p_{in}^c$ is reached. For more than two clusters, we also observe this bifurcation for the order parameter pair $\eta_{1,w-1}$, $\eta_{0,w}$. Second, the ground state energy remains on the level of $E^{Rnd}$ until $p_{in}>p_{in}^c$. Third, the ground state configuration has only random overlap with the planted partition, as seen from the plot of the accuracy, until $p_{in}>p_{in}^c$. This means that as long as $p_{in}<p_{in}^c$, the planted partition does not influence the ground state and is hence not detectable! The value of $p_{in}^c$=7/8 at which the planted solution starts to influence the ground state is smaller than the na\"{i}ve guess $p_{in}^{n}=2 E^{Rnd}/\langle k\rangle$=25/27, the value for which the planted solution starts to have an energy below $E^{Rnd}=25/18$.
\begin{figure*}[t]
\includegraphics[width=5.75cm]{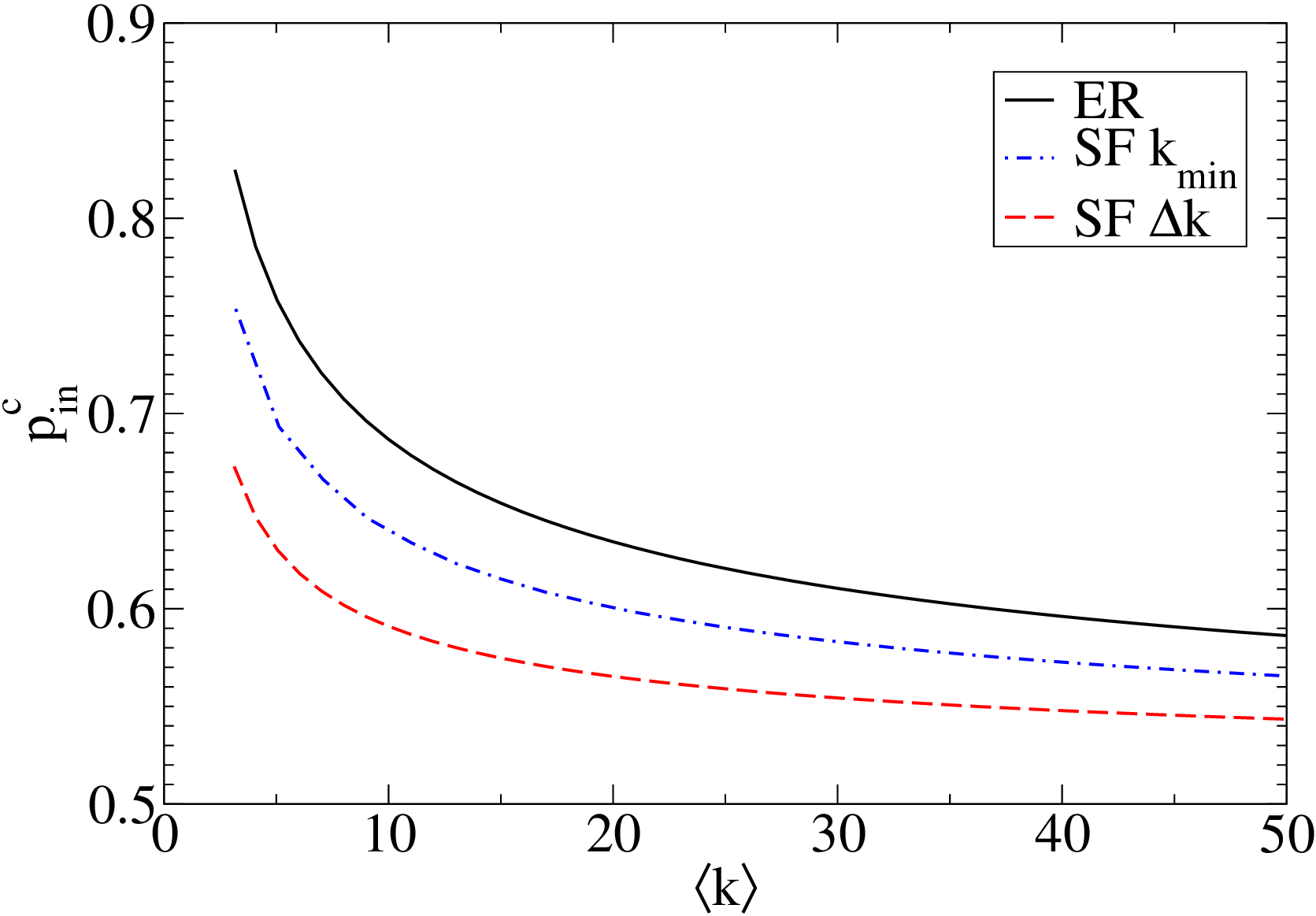}
\includegraphics[width=5.75cm]{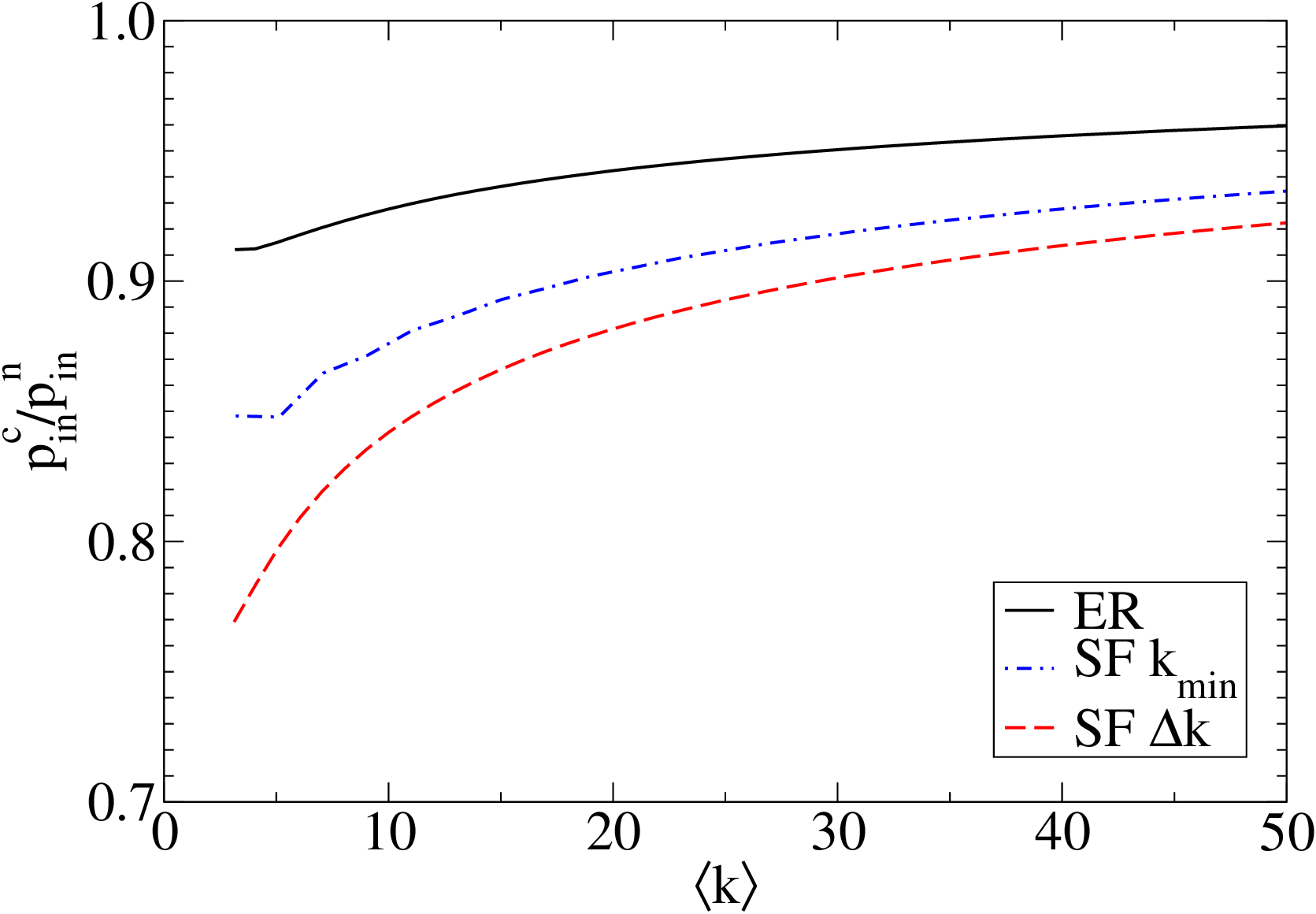}
\includegraphics[width=5.75cm]{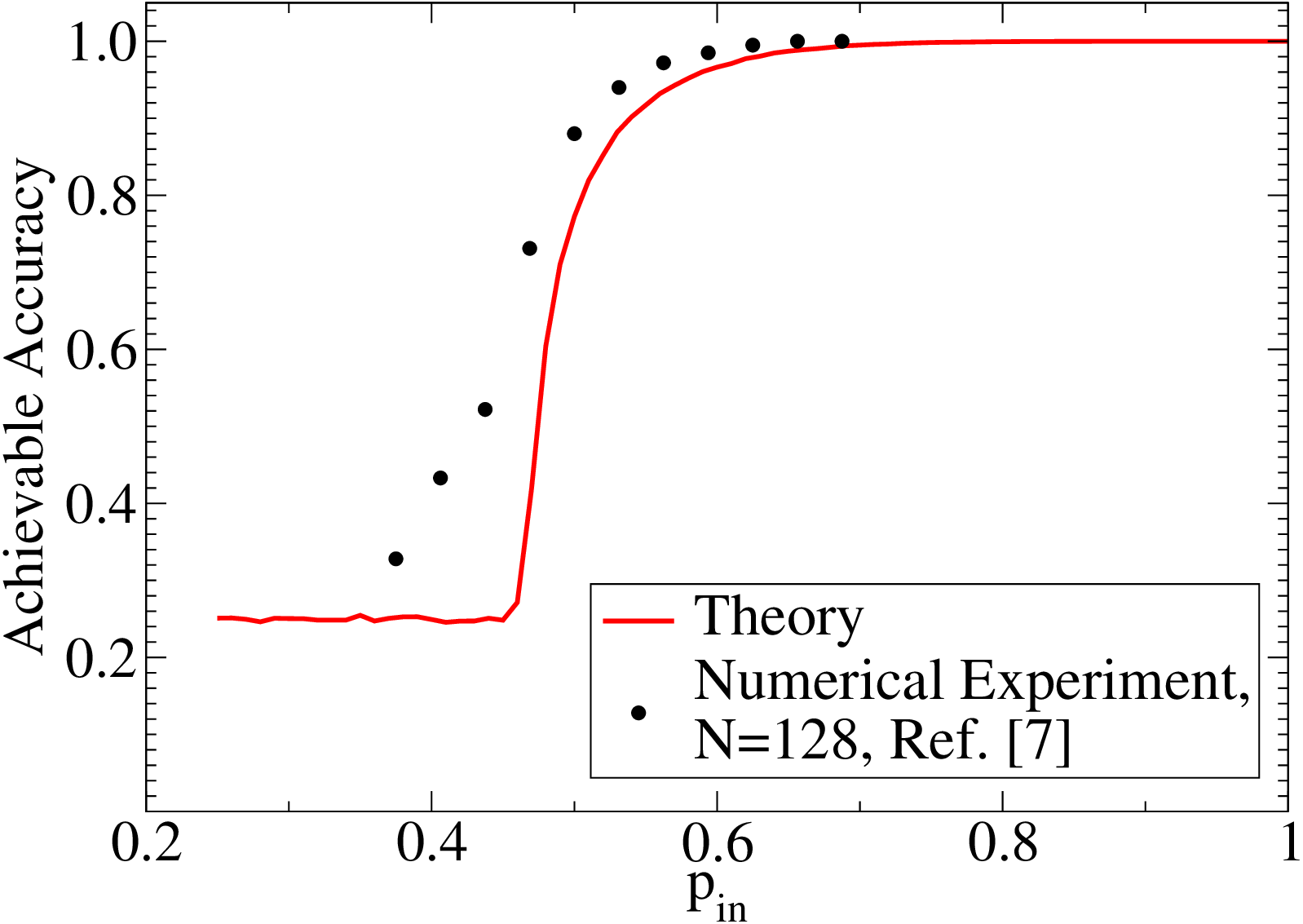}
\caption{\textbf{Left: }The critical value of $p_{in}$ beyond which the cluster structure starts to influence the ground state of the bisection problem, \ie below which clusters cannot be detected. We compare Erd\"os Renyi graphs (ER) with a Poissonian degree distribution $p(k)=e^{-\lambda}\lambda^k/k!$ and two types of scale free degree distributions. The first one being a stretched power law (SF $\Delta$k) of form $p(k)=(k+\Delta k)^{-\gamma}$ with $\Delta k\in[2,50]$, and the second (SF k$_{\mbox{\tiny min}}$) being of the form $p(k)=k^{-\gamma}$ with a varying minimum degree $k_{min}$ with $k_{\mbox{\tiny min}}\in [2,30]$. For both scale free distributions we choose $\gamma=3$. Since we are interested only in the behavior of the giant connected component, we set $p(k=0)=0$ in all cases. \textbf{Middle: } The ratio of $p_{in}^c$ and $p_{in}^n$, the na\"{\i}ve estimate for the transition point $p_{in}^n=2E^{Rnd}/\langle k\rangle$ which always overestimates $p_{in}^c$. \textbf{Right: } Achievable accuracy for the planted partition problem on ER graphs with $N\to\infty$, $\langle k\rangle=16$ and $4$ equal sized clusters and numerical results obtained from the best graph clustering algorithms on equivalent networks with $N=128$ nodes \cite{GuileraReview,GuimeraMeta}. We attribute the observed differences to finite size effects.}
\label{PinDifferentDegreeDist}
\end{figure*}

Let us now study how the critical value $p_{in}^c$ changes with the degree distribution. At the transition point, we can set $\eta_{10}=\eta_{01}+\delta\approx\eta_1$. Then we have $\eta^{in}_{10}=\eta_{10}-\delta p_{out}$ and $\eta^{in}_{01}=\eta_{10}-\delta p_{in}$. Inserting these expressions in (\ref{Eta11Full}-\ref{Eta10Full}) and expanding for small $\delta$  we arrive after at:
\begin{eqnarray}
(p_{in}^c-p_{out}^c)^{-1} & = &\sum_{n_0=0}\sum_{n_1>n_2}q(n_0+n_1+n_2)(n_1-n_2)\times\nonumber\\
& & \times\frac{(n_0+n_1+n_2)!}{n_0!n_1!n_2!}\eta_{1}^{n_1+n_2-1}\eta_{2}^{n_0}.
\label{PinCritical}
\end{eqnarray}
Here, we use with $\eta_1$ and $\eta_2$ the order parameters that we calculate for $p_{in}=1/2$ and that remain valid for all $p_{in}\leq p_{in}^c$. Again, expression (\ref{PinCritical}) is easily calculated for any degree distribution. In case of a Poissonian degree distribution $p(k)=e^{-\lambda}\lambda^k/k!$ with mean $\lambda$, we can simplify (\ref{PinCritical}) to
\begin{equation}
(p_{in}^c-p_{out}^c)^{-1}=\lambda\left(\eta_{2}+\frac{X_\lambda}{\eta_{1}}\right).
\end{equation}

Figure \ref{PinDifferentDegreeDist} shows the dependence of $p_{in}^c$ on the degree distribution. As a general feature $p_{in}^c$ decreases with increasing $\langle k\rangle$. However, the critical $p_{in}$ for distributions with fat tails is lower than for networks with a Poissonian degree distribution. Note the correspondence to the results in Ref. \cite{ReichardtPRER} on the cut-size of these graphs. The critical value of $p_{in}$ is smaller, \ie clusters are easier to detect, for networks with degree distributions which are harder to cut. Ref. \cite{ReichardtPRER} suggests a universal dependence of $E^{Rnd}$ on $\langle \sqrt{k}\rangle$ based on a replica calculation. Our calculations here support this result.  As the middle panel of figure \ref{PinDifferentDegreeDist} shows, in the limit of large $\langle k\rangle$ the na\"{i}ve estimate $p_{in}^c\approx p_{in}^n=2E^{Rnd}/\langle k\rangle$ provides a good, but conservative, approximation.

All the results described here analytically for two clusters can be obtained for more than two clusters by an efficient population dynamics algorithm which will be described elsewhere \cite{ReichardtLeonePrep}. As an example, the right panel shows the maximum theoretically attainable accuracy for a commonly used benchmark in graph clustering or community detection \cite{Girvan,GuileraReview}. While the actual benchmark uses networks with 128 nodes in 4 clusters and an average of 16 links per node, we calculate the accuracy for an infinitely large network with the same number of clusters and degree distribution. We recover the transition point and the upper part of the transition from the best available graph clustering algorithms \cite{GuimeraMeta,GuileraReview}.  Given the fact that the numerical experiments were obtained on a relatively small network and our theory applies to the thermodynamic limit, the aggreement between theory and experiment is remarkably good.  

In summary, we have shown that the sparsity of a network may limit the use unsupervised clustering methods may have. Even though cluster structure is present, it remains undetectable and hidden behind alternative solutions to the clustering problem that have zero correlation with the true solution. If we were to draw an analogy to unsupervised learning problems on multivariate data, we could say the average connectivity of a network plays the role of the ratio $\alpha$ between the number of data points and the dimensionality of a multivariate data set. The fundamental difference is that the average connectivity is not a free parameter in sparse networks and cannot be increased by adding more nodes to the network. Adding nodes to the network inevitably increases the dimensionality of the data. Thus we are dealing with a qualitatively different phenomenon. Our results may be valuable for the design of network clustering algorithms and their benchmarks as well as for a critical assessment of the amount of information that can be derived from unsupervised learning or data-mining on networks. 

We thank David Saad, Wolfgang Kinzel and Georg Reents for stimulating discussions. 
\bibliography{../BibTex_Citations}

\end{document}